\documentclass
[twocolumn,aps,prd,amsmath,amssymb,floatfix,superscriptaddress]
{revtex4-1}
\usepackage{silence} 
\usepackage[T1]{fontenc}             %
\usepackage{CJK}
\usepackage[dvips]{graphicx}         %
\usepackage{bm}                      
\usepackage{mathptmx}                
\usepackage{dcolumn}                 
\usepackage{multirow}                
\usepackage{xcolor}                  %
\usepackage{comment}                 %
\usepackage[
breaklinks,pdfstartview=FitH,CJKbookmarks=true,
bookmarksnumbered=true,bookmarksopen=true,pdfborder={0 0 1},
colorlinks=true,linkcolor=blue,urlcolor=blue,anchorcolor=blue,citecolor=blue
           ]{hyperref}

\begin{document}

\def\bea{\begin{eqnarray}} \def\eea{\end{eqnarray}}
\def\be{\begin{equation}} \def\ee{\end{equation}}
\def\bal#1\eal{\begin{align}#1\end{align}}
\def\ra{\rightarrow}
\def\arsinh{\mathop{\text{arsinh}}}
\def\al{\alpha}
\def\la{\Lambda}
\def\eps{\epsilon}
\def\gam{\gamma}
\def\om{\omega}
\def\ms{M_\odot}
\def\mmax{M_\text{max}}
\def\fm3{\;\text{fm}^{-3}}
\def\km{\;\text{km}}
\def\mev{\;\text{MeV}}
\def\gev{\;\text{GeV}}
\def\mfm{\;\text{MeVfm}^{-3}}

\long\def\OFF#1{}
\long\def\hj#1{\color{red}#1\color{black}}

\title{
Dark matter effects on the properties of hybrid neutron stars
}

\begin{CJK*}{UTF8}{gbsn}

\author{Jin-Biao Wei (魏金标)} \email{Contact author: weijb@cug.edu.cn}
\author{G. Wu (吴干)}
\author{H. Chen (陈欢)} \email{Contact author: huanchen@cug.edu.cn}
\affiliation{
School of Mathematics and Physics, China University of Geosciences,
Lumo Road 388, 430074 Wuhan, China}

\author{G. F. Burgio}
\author{H.-J. Schulze}
\affiliation{
INFN, Sezione di Catania,
Dipartimento di Fisica, Universit\'a di Catania,
Via Santa Sofia 64, 95123 Catania, Italy}

\begin{abstract}
We study the effects of dark matter on the properties of hybrid neutron stars,
in particular the influence on the mass-radius relation,
the value of the maximum mass,
and the hadron-quark phase transition.
To single out the equilibrium configurations
of dark-matter-admixed hybrid neutron stars (DHSs),
we also study their radial oscillations.
Both the stellar structure equations and the radial oscillation equations
are solved for the two-fluid system,
where the ordinary matter component and dark matter component
couple only through gravity.
For the ordinary matter components,
we adopt the Brueckner-Hartree-Fock method for nuclear matter,
and the Dyson-Schwinger or the field-correlator model for quark matter.
For the dark matter component,
we use a non-self-annihilating self-interacting fermionic model.
We find that the presence of dark matter in DHSs
leads to a decrease of the critical mass of the hadron-quark phase transition,
a related possible onset of quark matter in dark-matter accreting stars,
and a significant reduction of radial oscillation frequencies.
\end{abstract}

\maketitle
\end{CJK*}

\section{Introduction}

The origin and nature of dark matter (DM) represent one of the most profound
challenges in contemporary particle physics and cosmology
over the past decades \cite{Trimble87,Bergstrom00}.
A prevailing consensus holds that DM consists of yet undiscovered
elementary particles,
either within or beyond the Standard Model \cite{Bertone05,Feng10}.
First postulated by Zwicky in 1933 \cite{Zwicky09},
these particles constitute over 90\% of the Universe's matter content.
Their existence is inferred from observations across galactic
and supergalactic scales
\cite{Begeman91,Wittman00,Abdalla09,Abdalla10,Massey10,Scognamiglio26},
yet key properties
- such as mass and interactions with Standard Model particles -
remain undetermined
\cite{Henriques90,Ciarcelluti11,Guever14,Raj18}.

Experimental strategies to detect DM primarily revolve around three
complementary approaches:
direct detection, indirect detection, and production at particle colliders.
Direct detection seeks low-energy nuclear recoils from DM particles scattering
off target nuclei in underground detectors,
such as DAMA/LIBRA
\cite{Bernabei08,Bernabei10},
CRESST-I \cite{Bravin99},
and XENON \cite{Aprile12,Aprile18};
indirect detection targets annihilation or decay products
like gamma rays or positrons from dense DM regions
\cite{Abdalla09,Abdalla10,LeDelliou15},
whereas collider searches aim to produce DM particles
in high-energy proton collisions via missing transverse energy signatures,
as in SHiP at the Large Hadron Collider \cite{Deroeck24}.

Many particle physics scenarios suggest that compact stars
- white dwarfs and neutron stars (NSs) -
may also host a DM component \cite{Bramante24}.
Because of their large baryonic density,
the probability of interaction of DM with ordinary nuclear matter (NM)
could be large and DM capture would be increased.
Capture might take place during the star's lifetime
and therefore DM would accumulate steadily in the stellar interior
\cite{Bramante14,Stref17,Bell21},
leading to a DM-admixed compact object,
if nongravitational interactions between NM and DM exist
\cite{Lavallaz10,Lopes11,Bramante13,Bertoni13,Baryakhtar17,Bell21,Anzuini21}.
The efficiency of this process depends on several factors,
such as the local DM density,
the scattering cross-sections,
the production rate during core-collapse supernovae,
or capture in the progenitor star
\cite{Ciarcelluti11,Ellis18,Nelson19,Gleason22}.
Depending on the DM particle mass and self-interaction strength,
the dark component may be embedded into a NS compact core
or extend into a diffuse halo
\cite{Hippert22,Liu23,Liu24,Koehn24}.
In both cases,
the presence of DM changes dramatically the mass-radius relation,
mimicking the effect of softening  or stiffening the baryonic EOS
\cite{Bezares19,Ruter23}.
If annihilation reactions occur as well,
the compact objects may be heated and kinematical properties
such as linear and angular momentum may be significantly affected
\cite{Gonzales10,Angeles12,Herrero19,Garani21,Bose22,Fujiwara22,Coffey22}.

It is thus of great interest to theoretically analyze
DM-admixed NS (DNS) models.
The effects of DM on the properties of NM and NSs
have been intensely investigated in recent years,
typically combining a NM equation of state (EOS)
with a parameterized DM EOS within a general-relativistic two-fluid approach
\cite{Leung11,Li12,Tolos15,Maselli17,
Quddus20,Das20,Delpopolo20,Delpopolo20b,Das21,Yang21,Kain21,Leung22,Dengler22}.
A key insight is that DNS mass-radius relations cease to be unique,
instead varying systematically with DM fraction.
Given the profound uncertainty pertaining to both DM microphysics
and the high-density NM EOS,
theoretical inferences concerning either component remain inherently ambiguous.
For instance,
the observation of a very massive NS could plausibly arise
from a substantial DM core rather than an exceptionally stiff NM EOS alone.
Therefore it is hard to extract meaningful information by just comparing
the mass-radius relations predicted for purely baryonic stars and DNSs.
Consequently, robust theoretical diagnostics must be developed
to unambiguously discern the presence and abundance of DM in NSs.

\OFF{
A large number of works studied the possibility of DM capture by NSs
\cite{Lopes11,Bell21,Maity21,Anzuini21,Bose22}
and related phenomena like heating
\cite{Kouvaris08,Kouvaris10,Gonzales10,Bertoni13,Baryakhtar17,Raj18,
Garani21,Coffey22,Fujiwara22}
or internal black hole formation and collapse
\cite{Goldman89,Sandin09,Lavallaz10,Kouvaris12,McDermott12,
Bramante13,Bramante14,Bramante15,Ivanytskyi20}.

Recently, more quantitative studies have been performed.
For example, Ref.~\cite{Quddus20} indicated that the presence of DM in NSs
will soften the EOS and reduce the values of NS observables
like mass, radius, tidal deformability, and moment of inertia.
Ref.~\cite{Das20} investigated the DM effects
on the derived NM parameters like incompressibility, symmetry energy,
and higher-order derivatives like
slope parameter $L$, isovector incompressibility $K_\text{sym}$,
and skewness parameter $Q_\text{sym}$.
Ref.~\cite{Maselli17} studied the equilibrium structure
of rotating bosonic and fermionic dark stars
and indicated the existence of universal relations between the
moment of inertia $I$,
tidal deformability (Love number),
and quadrupole moment $Q$ ($I$-Love-$Q$ relations),
similar to those of normal NSs
\cite{Yagi13,Yagi13b,Yagi17}.
Ref.~\cite{Zhang20} combined a bosonic DM EOS with several NM EOSs
and examined the possibility of the LIGO/Virgo events
GW170817 \cite{Abbott17} and GW190425 \cite{Abbott20}
being realized by this DNS scenario.
Many recent works
\cite{Maselli17,Ellis18,ADas19,Nelson19,Quddus20,Husain21,Das21b,Das21c,
Das22,Leung22,Lourenco22,Karkevandi22,Dengler22,Hippert22}
focused on DM effects on the NS tidal deformability
and related observables \cite{Das21d},
which are directly accessible by current and future GW detectors.
The impact of DM on the pulsar x-ray profile \cite{Miao22}
and on NS cooling processes \cite{Bhat20,Kumar22}
were also studied recently.
}

Recently we have investigated the global properties of DNSs
within the Brueckner-Hartree-Fock (BHF) method for the nuclear EOS
and generic self-interacting bosonic and fermionic DM models
for the respective EOSs,
trying to identify precise signals for the presence of DM in NSs
\cite{Liu23,Liu24,Routaray25}.
Cooling properties have also been analyzed,
to highlight the crucial role of direct Urca processes and superfluidity gaps,
identifying possible signals of a large DM content \cite{Zhou25}.

The present study is a further extension of those works.
Due to the large NS central density,
asymptotic freedom enables quark deconfinement
and therefore quarks may become more energetically favorable than baryons.
This leads to the possibility that the core of a NS could be made up
of deconfined quark matter (QM) surrounded by a hadronic layer,
thus forming a hybrid star (HS) \cite{gle}.
We investigate here the effects of DM on an eventual
hadron-quark first-order phase transition in NS matter,
and whether this invalidates conclusions made so far.

In particular,
we study how the presence of a DM component influences the mass-radius relation,
the value of the maximum mass,
and the main features of a hadron-quark phase transition.
For this purpose we consider a hybrid-star EOS
built via a Gibbs construction
from a BHF theoretical approach for nucleons only,
and two models of QM, viz.,
a Dyson-Schwinger based model (DSM) for $uds$ QM,
as already studied in \cite{Roberts00,Chen08,Chen12},
and the Field Correlator Model (FCM)
\cite{Simonov07a,Simonov07b,Nefediev09}.
For the DM EOS we use a non-self annihilating self-interacting Fermi gas
with particle mass $\mu=1\gev$.
These two fluids, NM and DM, interact only gravitationally.

We emphasize that a scenario of very large DM fractions
$f\sim{\cal{O}}(1)$
that we consider here
is very speculative and requires exotic formation mechanisms of DNSs like
dark conversion of the neutron to scalar DM,
neutron bremsstrahlung of scalar DM,
mirror DM,
self-interaction of condensate DM
\cite{Foot04,Sandin09,XYLi12,Perez22,Garcia22},
since the standard DM capture mechanism of diluted DM in the Universe
\cite{Ellis18}
is only able to provide DM fractions of about $f\sim10^{-10}$
for NSs in the galactic centers.
Therefore even DM fractions of a few percent
that have been assumed in many articles,
are as extreme as 10 or 90 percent,
and demand nonstandard explanations.
With this in mind we perform this study as continuation of
\cite{Liu23,Liu24,Zhou25},
where we analyzed the global properties of $f=0,...,1$ DNSs.

This article is organized as follows.
In the next section,
the EOSs of ordinary NM and DM used in this work are briefly described.
The detailed calculations and discussion are presented in Sec.~\ref{s:res},
and a summary is given in Sec.~\ref{s:sum}.
Throughout the paper, we use natural units $G=c=\hbar=1$.

\section{Formalism}

\subsection{EOS for ordinary matter}

For the ordinary matter,
we employ different theoretical approaches for the nuclear and quark phases,
which are joined by the Gibbs construction method
for the hadron-quark phase transition.

To describe the NM phase,
we use the microscopic BHF approach
\cite{Baldo99,Baldo12,Burgio21a},
where the only required input quantity
is the bare nucleon-nucleon interaction $V$.
In the present work,
we employ the Argonne $V_{18}$ potential \cite{Wiringa95},
supplemented by a consistent meson-exchange three-nucleon force,
which allows to reproduce correctly the NM saturation point
\cite{Grange89,Zuo02,Li08a,Li08b} 
and other properties of NM around saturation
\cite{Wei20,Burgio21b}.
Furthermore,
the value of the NS maximum mass $\mmax=2.34\ms$ for the V18 EOS
is compatible with the current observational lower limits
\cite{Antoniadis13,Arzoumanian18,Fonseca21};
in particular the recent data $M=2.35\pm0.17\ms$ for PSR J0952-0607
\cite{Romani22}.
The radii $R_{1.4}=12.32\km$ and
$R_{2.0}=11.93\km$
are compatible with the most recent combined GW170817+NICER analysis
\cite{Rutherford24},
$R_{1.4}=12.28^{+0.50}_{-0.76}\km$ 
and
$R_{2.0}=12.33^{+0.70}_{-1.34}\km$, 
and the tidal deformability $\Lambda_{1.4}=430$
agrees with the observed value $\approx70-580$
\cite{Abbott17,Abbott18,Burgio18,Wei19}.

\OFF{
The most massive cold NSs observed so far are
PSR~J1614-2230 ($M=1.908\pm0.016\ms$) \cite{Arzoumanian18},
PSR~J0348+0432 ($M=2.01\pm0.04\ms$) \cite{Antoniadis13}, and
PSR~J0740+6620  2.14-0.09+0.10 \cite{Cromartie20}, OLD!
PSR~J0740+6620 ($M=2.08\pm0.07\ms$) \cite{Fonseca21},
PSR~J0952-0607 ($M=2.35\pm0.17\ms$) \cite{Romani22}

PSR~J0740+6620 with measured radius
$R(2.08\pm0.07\ms) = 13.7^{+2.6}_{-1.5}\km$ \cite{Riley21}
$R(2.072^{+0.067}_{-0.066}\ms) = 12.39^{+1.30}_{-0.98}\km$ \cite{Miller21}
$R(2.073\pm0.069\ms) = 12.49-0.88+1.28 \cite{Salmi24} 11.6,12.5,13.8
12.76-1.02+1.49 or 12.92-1.13+2.09 \cite{Dittmann24}

We also mention the combined estimates of the mass and radius
of the isolated pulsar J0030+0451 observed recently by NICER,
$M=1.44^{+0.15}_{-0.14}\ms$ and $R=13.02^{+1.24}_{-1.06}\km$
\cite{Miller19,Riley19}
$M=1.36^{+0.15}_{-0.16}\ms$ and $R=12.71^{+1.14}_{-1.19}\km$ 11.5,12.7,13.8
\cite{Miller21,Riley21}
$M=1.40^{+0.13}_{-0.12}\ms$ and $R=11.71^{+0.88}_{-0.83}\km$ 10.9,11.7,12.6
$M=1.70^{+0.18}_{-0.19}\ms$ and $R=14.44^{+0.88}_{-1.05}\km$
\cite{Vinciguerra24}

and in particular the result of the combined GW170817+NICER analysis
\cite{Abbott17,Abbott18},
$R_{2.08}=12.35\pm0.75\km$ \cite{Miller21}, and
$R_{1.4}=
$12.45\pm0.65$ km \cite{Miller21}, 
$11.94^{+0.76}_{-0.87}$ km \cite{Pang21}, and
$12.33^{+0.76}_{-0.81}$ km or
$12.18^{+0.56}_{-0.79}$ km \cite{Raaijmakers21}
$12.28^{+0.50}_{-0.76}$ km \cite{Rutherford24}PP
$12.01^{+0.56}_{-0.75}$ km \cite{Rutherford24}CS
$R_{2.0}=12.33^{+0.70}_{-1.34}$ km \cite{Rutherford24}PP
$R_{2.0}=11.55^{+0.94}_{-1.09}$ km \cite{Rutherford24}CS
}

To describe the QM phase, we employ two different quark models:

(i) Dyson-Schwinger quark model (DSM)
\cite{Roberts00,Chen08}:
The DSM is a continuum approach to QCD that can address
both confinement and dynamical chiral symmetry breaking.
It has been successfully applied to hadron physics in vacuum
\cite{Roberts00,Roberts07}
and extended to finite chemical potential and temperature
\cite{Chen08,Chen12,Chen16}.
Two important parameters are the interaction strength $\al_\text{DS}$
and the bag constant $B_\text{DS}$,
and as in \cite{Sun21,Zheng23}
we choose in this work the values
$B_\text{DS}=38\mfm$ for $\al_\text{DS}=1$ and
$B_\text{DS}=138\mfm$ for $\al_\text{DS}=1.5$,
which both yield a maximum mass $2.1\ms$ of HSs.

(ii) Field correlator method (FCM)
\cite{Simonov07a,Simonov07b,Nefediev09}:
The FCM is a nonperturbative approach to QCD that incorporates the effects
of confinement and chiral symmetry breaking through field correlators.
It has been applied to study the EOS of QM in NSs
\cite{Baldo08,Bombaci12,Plumari13,Alford15,Burgio15,Wei20}.
The FCM EOS depends on two parameters:
the gluon condensate $G_2$ and
the large-distance static $q\bar{q}$ potential $V_1$.
We choose $G_2=0.006\gev^4$ and $V_1=142\mev$ in this work,
to achieve a maximum mass of $2.1\ms$, as for the DSM EOS.

Once the EOSs of NM and QM are specified,
we are able to construct the EOS of HS matter
by performing the Gibbs phase transition,
according to which there is a mixed phase
where the hadron and quark phases coexist,
and both phases are in equilibrium with each other \cite{gle}.
This can be expressed as
\be
 \mu_i = b_i \mu_B - q_i \mu_e \ , \quad p_H = p_Q = p_M \:,
\ee
where $b_i$ and $q_i$ denote baryon number and charge of the particle species
$i=n,p,u,d,s,e,\mu$ in the mixed phase.
These equations are solved together with the global charge neutrality condition
\be
 \chi\rho_c^Q + (1-\chi)\rho_c^H = 0 \:,
\ee
where $\rho_c^Q$ and $\rho_c^H$ are the charge densities
of QM and NM,
and $\chi$ is the volume fraction occupied by QM in the mixed phase.
From these equations,
one can obtain the energy density and the baryon density of the mixed phase as
\bea
 \eps_M &=& \chi\eps_Q + (1-\chi)\eps_H \:,
\\
 \rho_M &=& \chi\rho_Q + (1-\chi)\rho_H \:.
\eea

\begin{figure}[t]
\vskip-2mm
\centerline{\hskip0mm\includegraphics[scale=0.62]{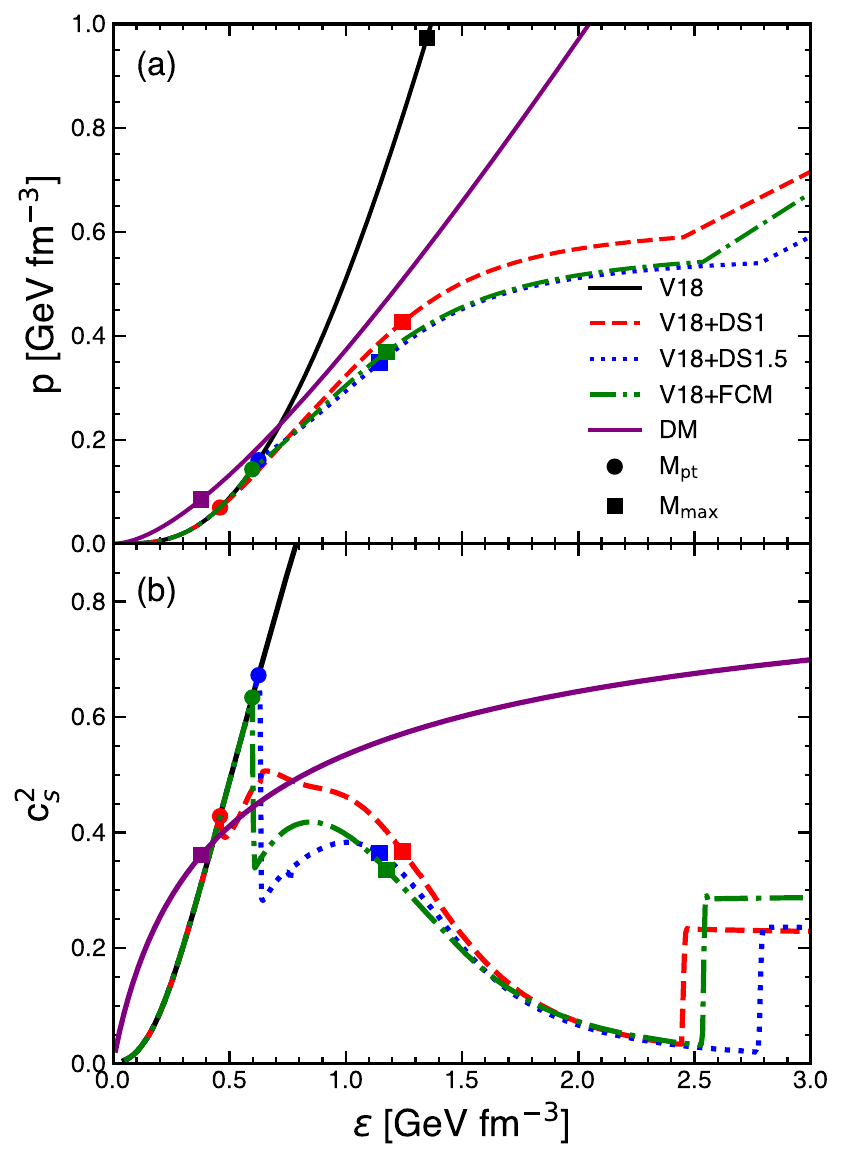}}
\vskip-4mm
\caption{
(a)
Stellar EOSs of
nuclear matter (solid black curve),
hybrid matter with Gibbs phase transition construction (broken colored),
and for DM (solid lilac);
(b)
Squared speed of sound as function of energy density $\eps$
for the selected EOSs.
Markers indicate the onset of QM
and the $\mmax$ configurations of the corresponding stars.
}
\label{f:eos}
\end{figure}

In Fig.~\ref{f:eos}(a) we compare the resulting EOSs of HSs
with those of pure NM and of pure DM, 
introduced later.
One notes an early onset of the extended mixed phase,
which considerably smoothes the hybrid EOS.
In fact pure QM phases are never reached in HSs up to their
maximum masses.
The squared speed of sound $c_s^2=\partial p/\partial \eps$
is an important quantity that influences in particular the stellar oscillations,
as discussed later.
It is displayed in Fig.~\ref{f:eos}(b) for the various EOSs.
One notes results in accordance with the EOSs shown in panel (a),
namely the mixed phase strongly reduces $c_s^2$
compared to the nucleonic (and DM) EOS.

\begin{figure}[t]
\vskip-10mm
\centerline{\hskip0mm\includegraphics[scale=0.64]{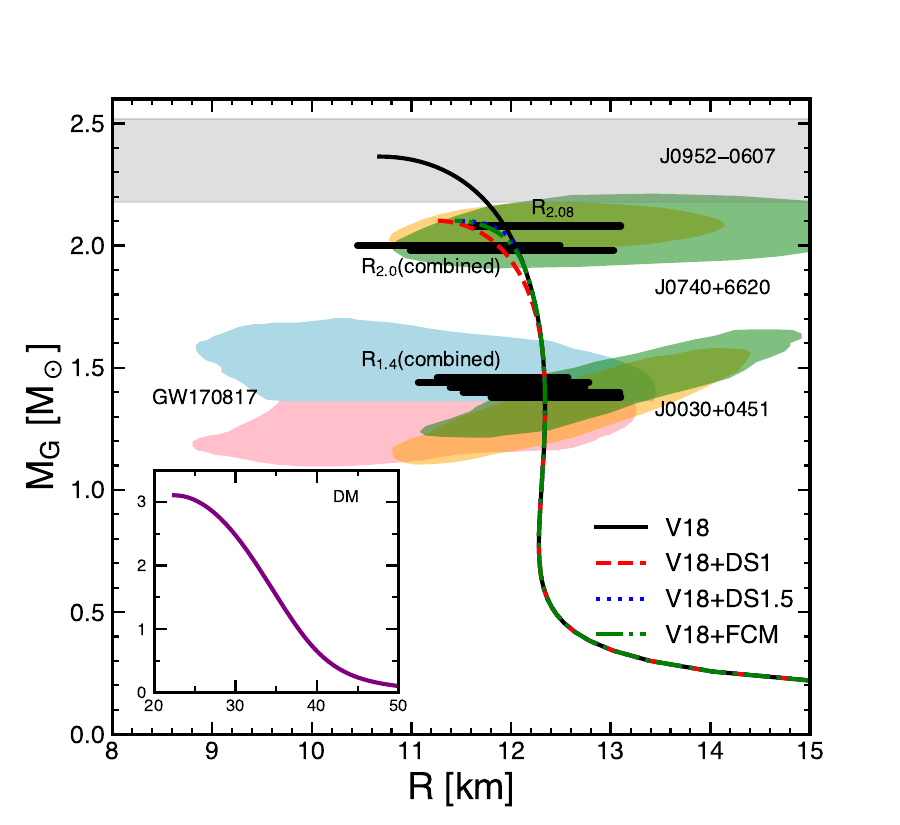}}
\vskip-4mm
\caption{
Mass-radius relation for the various EOSs.
Observational constraints on masses
\cite{Fonseca21,Romani22}
and radii $R_{1.4}$, $R_{2.0}$, and $R_{2.08}$ from NICER
\cite{Miller21,Rutherford24}
(horizontal bars)
are included.
}
\label{f:mr}
\end{figure}

Fig.~\ref{f:mr} shows the mass-radius relations for the various EOSs,
together with current observational constraints.
According to the most recent and accurate
observations of massive NSs we have
PSR J1614-2230 ($M=1.908\pm0.016\ms$) \cite{Arzoumanian18},
PSR J0348+0432 ($M=2.01\pm0.04\ms$) \cite{Antoniades13},
PSR~J0740+6620 ($M=2.08\pm0.07\ms$) \cite{Fonseca21}, and
PSR J0952-0607 ($M=2.35\pm0.17\ms$) \cite{Romani22}.
Those imply that the maximum gravitational mass of NSs lies above $2\ms$.
The recent NICER data also give the combined mass and radius of
PSR J0030+0451 with
$R(1.44^{+0.15}_{-0.14}\ms)=13.02^{+1.24}_{-1.06}\km$
\cite{Miller19,Riley19},
$R(1.36^{+0.15}_{-0.16}\ms)=12.71^{+1.14}_{-1.19}\km$
\cite{Miller21,Riley21},
$R(1.40^{+0.13}_{-0.12}\ms)=11.71^{+0.88}_{-0.83}\km$
\cite{Vinciguerra24},
and for PSR J0740+6620 with
$R(2.08\pm0.07\ms) = 13.7^{+2.6}_{-1.5}\km$ \cite{Riley21},
$R(2.072^{+0.067}_{-0.066}\ms) = 12.39^{+1.30}_{-0.98}\km$ \cite{Miller21},
$R(2.073\pm0.069\ms) = 12.49^{+1.28}_{-0.88}$ \cite{Salmi24}.
Finally,
the most recent combined GW170817+NICER analysis
\cite{Rutherford24} yields
$R_{1.4}=12.28^{+0.50}_{-0.76}\km$ 
and
$R_{2.0}=12.33^{+0.70}_{-1.34}\km$ or 
$R_{2.0}=11.55^{+0.94}_{-1.09}\km$. 

As stated before,
the nucleonic V18 EOS is fully compatible with all current constraints,
whereas all hybrid EOSs are unable to reach
the observed mass value of PSR J0952-0607,
due to the reductions of their maximum mass to $2.1\ms$.
The pure DM EOS shown for comparison
features a $M(R)$ relation of the same order of magnitude as the nucleonic EOS,
for the mass parameters $\mu=1\gev$ considered in this work.

\subsection{EOS for dark matter}

Concerning the DM EOS,
considering the quasi complete lack of knowledge regarding the nature of DM,
in the present work we employ as a representative choice the frequently-used
\cite{Narain06,Tulin13,Kouvaris15}
DM model of fermions with mass $\mu$,
self-interacting via a repulsive Yukawa potential
\bal
 V(r) &= \al \frac{e^{-m r}}{r} \:
\eal
with coupling constant $\al$ and mediator mass $m$.
It allows to study in a simple and transparent way
the qualitative features of DM in NSs.
Detailed discussions of DM effects on pure NSs
have been given in our previous works \cite{Liu24,Routaray25,Zhou25}.

Currently, there are basically no constraints on the DM particle mass
\cite{Kouvaris11,Bramante13}. 
However, it turns out that only masses of the order of the nucleon mass
yield sizeable observable effects on typical NS observables by admixed DM
\cite{Liu24}.
For simplicity, we choose in this work $\mu=1\gev$
as a representative value,
which allows to examine the qualitative impact of DM on NS properties.
For more complete studies including the $\mu$ degree of freedom we refer
to \cite{Liu24,Routaray25,Zhou25}.

Following \cite{Narain06} we write pressure and energy density
of the resulting DM EOS as
\bal
 p_D    &= \frac{\mu^4}{8\pi^2}
 \Big[ x\sqrt{1+x^2}(2x^2\!/3-1) + \arsinh{(x)} \Big]
 + \delta \:,
\\
 \eps_D &= \frac{\mu^4}{8\pi^2}
 \Big[ x\sqrt{1+x^2}(2x^2+1) - \arsinh{(x)} \Big]
 + \delta \:,
\eal
where
\be
 x = \frac{k_F}{\mu} = \frac{(3\pi^2n)^{1/3}}{\mu}
\ee
is the dimensionless kinetic parameter with the DM-particle
Fermi momentum $k_F$ and density $n$,
and the self-interaction term is written as
\bal
 \delta =
 \frac{2}{9\pi^3} \frac{\al \mu^6}{m^2} x^6 &\equiv
  \mu^4 \Big(\frac{y}{3\pi^2}\Big)^2 x^6
 = \Big(\frac{yn}{\mu}\Big)^2 \:,
\label{e:y}
\eal
introducing the interaction parameter
$y^2 = 2\pi\al\mu^2\!/m^2$.

Within this model,
$\mu$ and $y$ are not independent free parameters, but constrained
by limits imposed on the DM self-interaction cross section $\sigma$
through observation of the interaction of galaxies
in different colliding galaxy clusters
\cite{Markevitch04,Kaplinghat16,Sagunski21,Loeb22},
\be
 \sigma/\mu \sim 0.1-10 \;\text{cm}^2\!/\text{g} \:.
\ee
In \cite{Tulin13,Kouvaris15,Maselli17}
it has been shown that the Born approximation
\bal
 \sigma_\text{Born} &= \frac{4\pi\al^2}{m^4}\mu^2 = \frac{y^4}{\pi\mu^2}
\eal
is very accurate for $\mu\lesssim1\gev$
and in any case remains valid in the limit $\al\ra0$ for larger masses.
We therefore employ here this approximation,
choosing for simplicity the fixed constraint
\be
 \sigma/\mu = 1 \;\text{cm}^2\!/\text{g} = 4560/\text{GeV}^3 \:,
\ee
which leads to
\bal
 y \simeq 10.94\,\mu_1^{3/4}, \qquad \mu_1 \equiv \mu/(1~\mathrm{GeV}).
\eal
After this choice, the DM EOS depends only on the single parameter $\mu$.
As stated before,
we choose $\mu=1\gev$ in this work and thus $y=10.94$.
The resulting DM EOS and the corresponding mass-radius relation
are also shown in Figs.~\ref{f:eos} and \ref{f:mr}, respectively.

\subsection{Hydrostatic configuration}

The stable configurations of DHSs
in the spherical metric
\be\label{e:ds2}
 ds^2 = e^{\nu(r)}dt^2 - e^{\lambda(r)}dr^2 -
 r^2(d\theta^2+\sin^2\!\theta d\varphi^2) \:
\ee
are obtained from a two-fluid version of the TOV equations
\cite{Kodama72,Comer99,Sandin09}:
\bal
 p_D' &= -q_D\nu'\!/2 \:,
\label{e:tovd}
\\
 p_N' &= -q_N\nu'\!/2 \:,
\label{e:tovn}
\\
 m' &= 4\pi r^2 \eps \:,
\label{e:tovm}
\\
 {m_B}' &= 4\pi r^2 \rho_B m_n e^\lambda \:,
\\
 \nu'\!/2 &=
 \frac{m + 4\pi r^3p}{r^2} e^{\lambda}\:,
\label{e:tovnu}
\eal
where $r$ is the radial coordinate from the center of the star,
the prime denotes radial derivative here and in the following,
$q_i \equiv \eps_i+p_i$,
$\nu$
and $e^\lambda=1/({1-2m/r})$ are metric functions,
and
$p=p_N+p_D$,
$\eps=\eps_N+\eps_D$,
$m=m_N+m_D$,
$m_B$
are the total pressure, total energy density,
and enclosed gravitational and baryonic mass, respectively.
The total gravitational mass of the DNS is
\be
 M_G = m_N(R_N) + m_D(R_D) \:,
\ee
where the stellar radii $R_N$ and $R_D$
are defined by the vanishing of the respective pressures.
There are thus in general two scenarios:
DM-core ($R_D<R_N$) or DM-halo ($R_D>R_N$) stars.

\subsection{Radial oscillations and stability}

As for the two-fluid TOV equations,
we assume that ordinary matter and DM interact only through gravity.
Thus also the stellar pulsation equations can be separated
into those of ordinary matter and DM.
However, deriving the equations for the radial oscillations of a two-fluid star
is still more complicated than in the one-fluid case.
This derivation has been done in several works
\cite{Comer99,Kain21,Gleason22,Caballero24,Zhen24},
and we follow the methodology of \cite{Kain21}.
Here we just present a brief introduction
and refer the reader to \cite{Kain21,Gleason22} for more details.

To study the radial oscillations,
one introduces independent Lagrangian radial displacements
with the common angular frequency $\om$ of the oscillation mode,
\be
 \xi_i(t,r) = \xi_i(r) e^{i\om t} \:,\quad  i=N,D \:,
\ee
and the related variables
\be
 \zeta_i(r) \equiv r^2 e^{-\nu(r)/2} \xi_i(r) \:.
\ee
The idea is to combine the Einstein field equations and equations of motion
such as to obtain a system of coupled pulsation equations
in terms of $\zeta_i$ and their derivatives $\zeta_i'$, 
which can be written as
\begin{widetext}
\bal
 & \left(\hat\Pi_i \hat\zeta_i'\right)'
 + \left( \hat{Q}_i + \hat\om^2 W_i \right) \hat\zeta_i
 + \hat{R} \bigg[
   \left(\frac{q_i}{r} - p_i'\right) \sum_j q_j \hat\zeta_j
   + \frac{r^2 q_i}{\hat\sigma^2 H} \sum_j \hat\Pi_j \hat\zeta_j'
           \bigg]
 =\; \hat{S}_i q_{\bar{i}} \Delta\hat\zeta_i
 + \hat{R} \gam_i p_i \left( q_{\bar{i}} \Delta\hat\zeta_i \right)'
 + \frac{r^2 q_i \hat{R}^2}{\hat\sigma^2 H}
 \sum_j \gam_j p_j q_{\bar{j}} \Delta\hat\zeta_j
\:,
\label{e:osc}
\eal
\end{widetext}
where
$\Delta\hat\zeta_i  
\equiv \hat\zeta_{\bar{i}} - \hat\zeta_i$
($\bar{i}$ is the complement of $i$),
and the adiabatic index
$\gam_i \equiv ({q_i}/{p_i}) ({\partial p_i}/{\partial\eps_i})$
for each fluid component.
The coefficients
$\hat\Pi_i$, $\hat{Q}_i$, $W_i$, $\hat{R}$, and $\hat{S}_i$
are given by
\bal
 \hat\Pi_i & = \gam_i p_i \hat\sigma^2 H / r^2 \:,
\\
 \hat{Q}_i & = -\frac{\hat\sigma^2}{r^2}
 \left[ \frac{3H p_i'}{r} + {8\pi p q_i}
 + \left( {4\pi r} \eps - \frac{m}{r^2} \right)
 \left( \frac{q_i}{r} - p_i' \right)
 \right] \:,
\\
 W_i &= q_i / (r^2 H) \:,
\\
 \hat{R} &= 4\pi\hat\sigma^2 \!/ r \:,
\\
 \hat{S}_i &= \hat{R}\left[
 \left(\gam_i-1\right) p_i' + \gam_i' p_i + \gam_i p_i
 \left( \frac{8\pi r}{H} q - \frac{1}{r}\right)
 \right] \:.
\eal
All quantities distributed in the star,
such as the energy density $\eps_i$,
the pressure $p_i$,
and the metric functions
$H \equiv 1 - 2m/r = e^{-\lambda}$ and
$\sigma \equiv e^{\nu/2}/\sqrt{H} = e^{(\nu+\lambda)/2}$
are obtained by solving the TOV equations.
Note that the quantities with a hat have been scaled by powers of
the central value of $\sigma$,
$\sigma_c=\sigma(r=0)$,
like
$\hat\zeta_i\equiv\zeta_i\sigma_c^2$,
$\hat\om\equiv\om/\sigma_c$, etc.

\begin{figure*}[t]
\vskip-1mm
\centerline{\hskip0mm\includegraphics[scale=0.54]{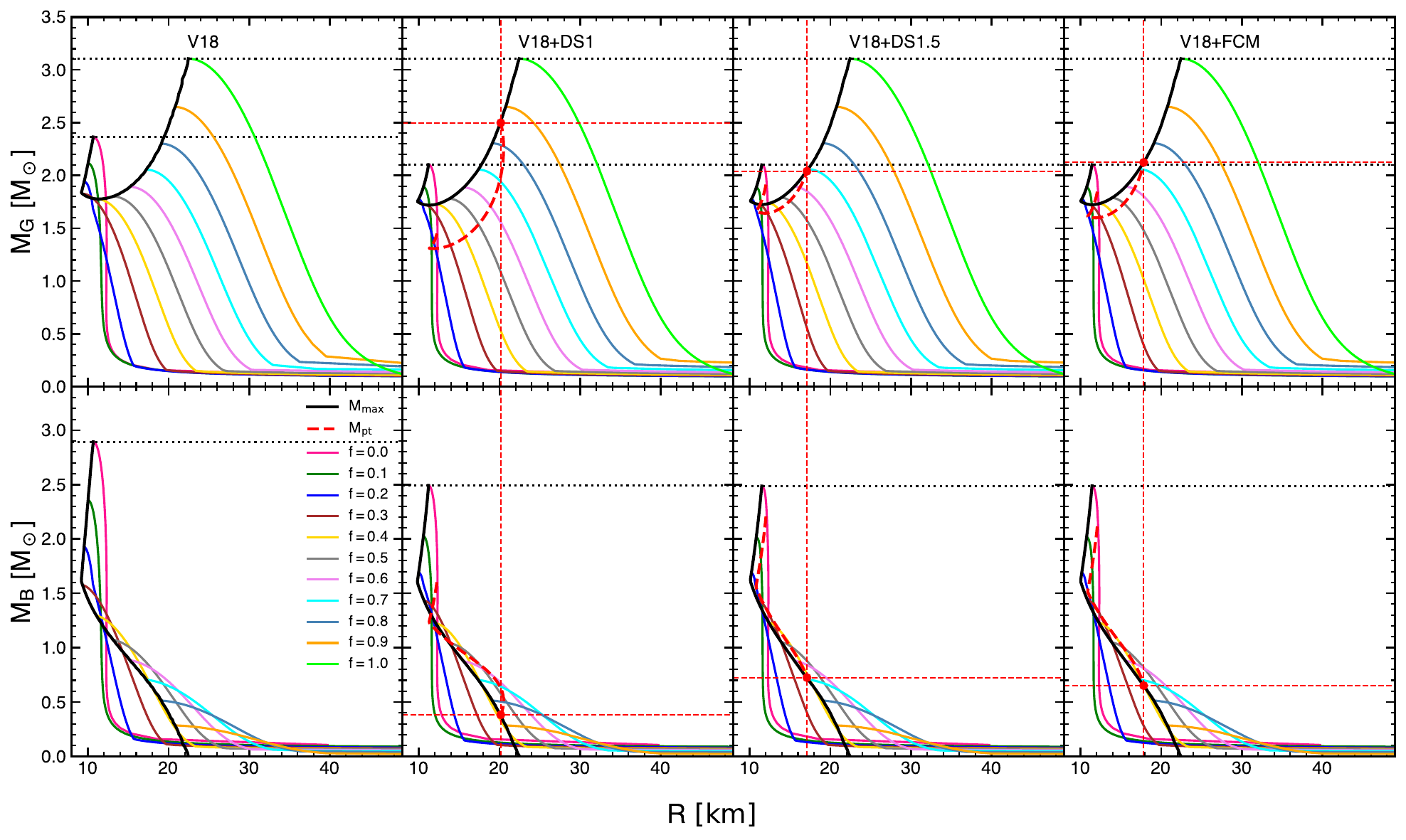}}
\vskip-4mm
\caption{
DHS mass $(M_G,M_B)$ vs radius $R=\max(R_N,R_D)$ relations
for different DM fractions
$f=M_D/M=0,0.1,0.2,...1$,
for various EOSs.
The solid black curves indicate the sequence of maximum masses
with varying $f$.
The broken red curves indicate the critical configurations
of the hadron-quark phase transition onset.
Dotted black horizontal lines indicate $\mmax$
of pure nucleonic and dark stars.
Markers and dashed red lines indicate the range of possible HSs.
See text for further details.
}
\label{f:mrf}
\end{figure*}

\begin{figure*}[t]
\vskip-1mm
\centerline{\hskip0mm\includegraphics[scale=0.54]{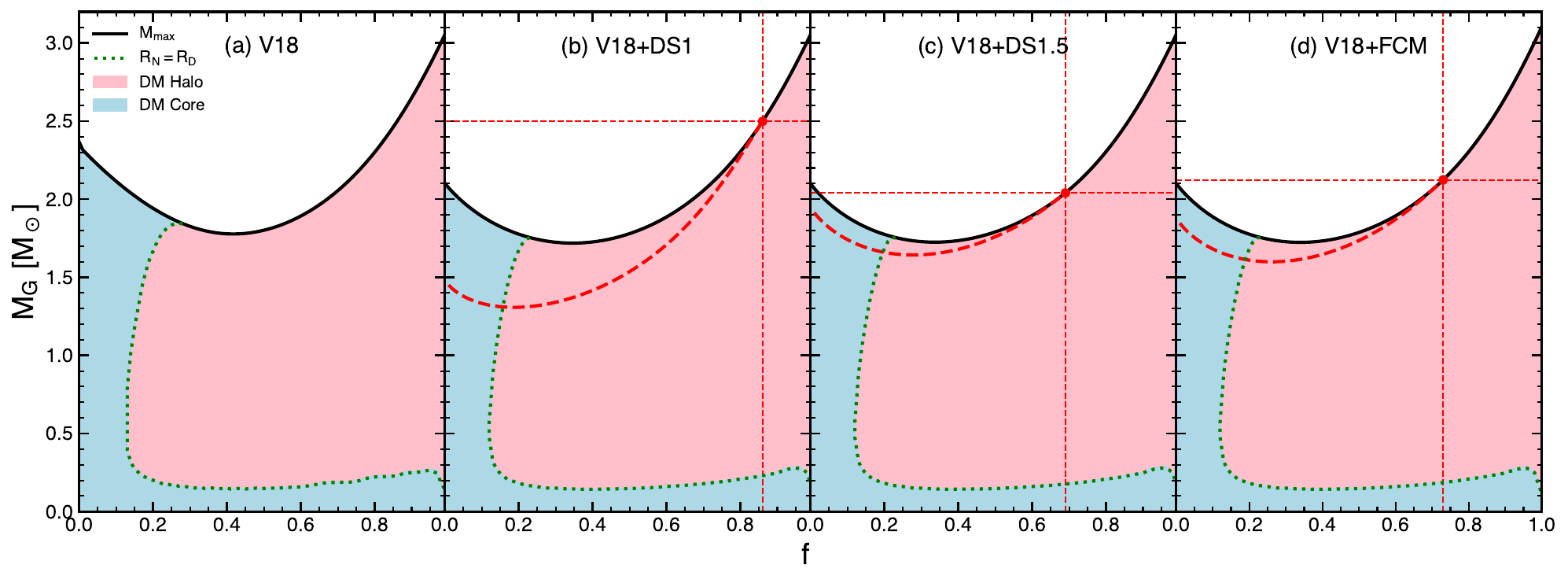}}
\vskip-4mm
\caption{
The stable DHS configurations in the $(f,M_G)$ plane
for the different EOSs.
The $\mmax$ and $M_\text{pt}$ curves correspond to those in Fig.~\ref{f:mrf}.
The DM-core and DM-halo domains are emphasized,
separated by the $R_N=R_D$ dotted green curves.
}
\label{f:mf}
\end{figure*}

\begin{figure}[t]
\vskip-1mm
\centerline{\hskip0mm\includegraphics[scale=0.57,angle=0]{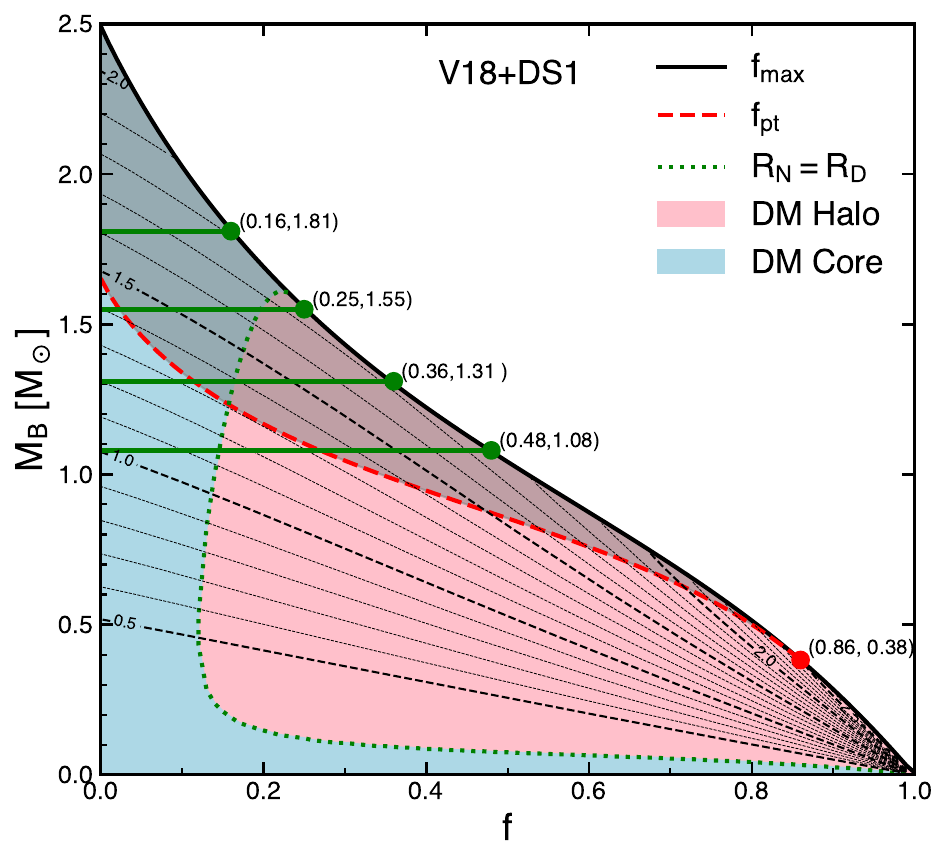}}
\vskip-4mm
\caption{
The allowed domain of DHSs with the EOS V18+DS1 in the $(f,M_B)$ plane.
Notation is as in Fig.~\ref{f:mf}.
Contours of gravitational mass $M_G$ are shown by dashed lines.
Horizontal green lines indicate  fixed-$M_B$ configurations
with initial $M_G(f=0)=1.0,1.2,1.4,1.6\,\ms$.
}
\label{f:mbf}
\end{figure}

\begin{figure}[t]
\vskip-1mm
\centerline{\includegraphics[scale=0.54]{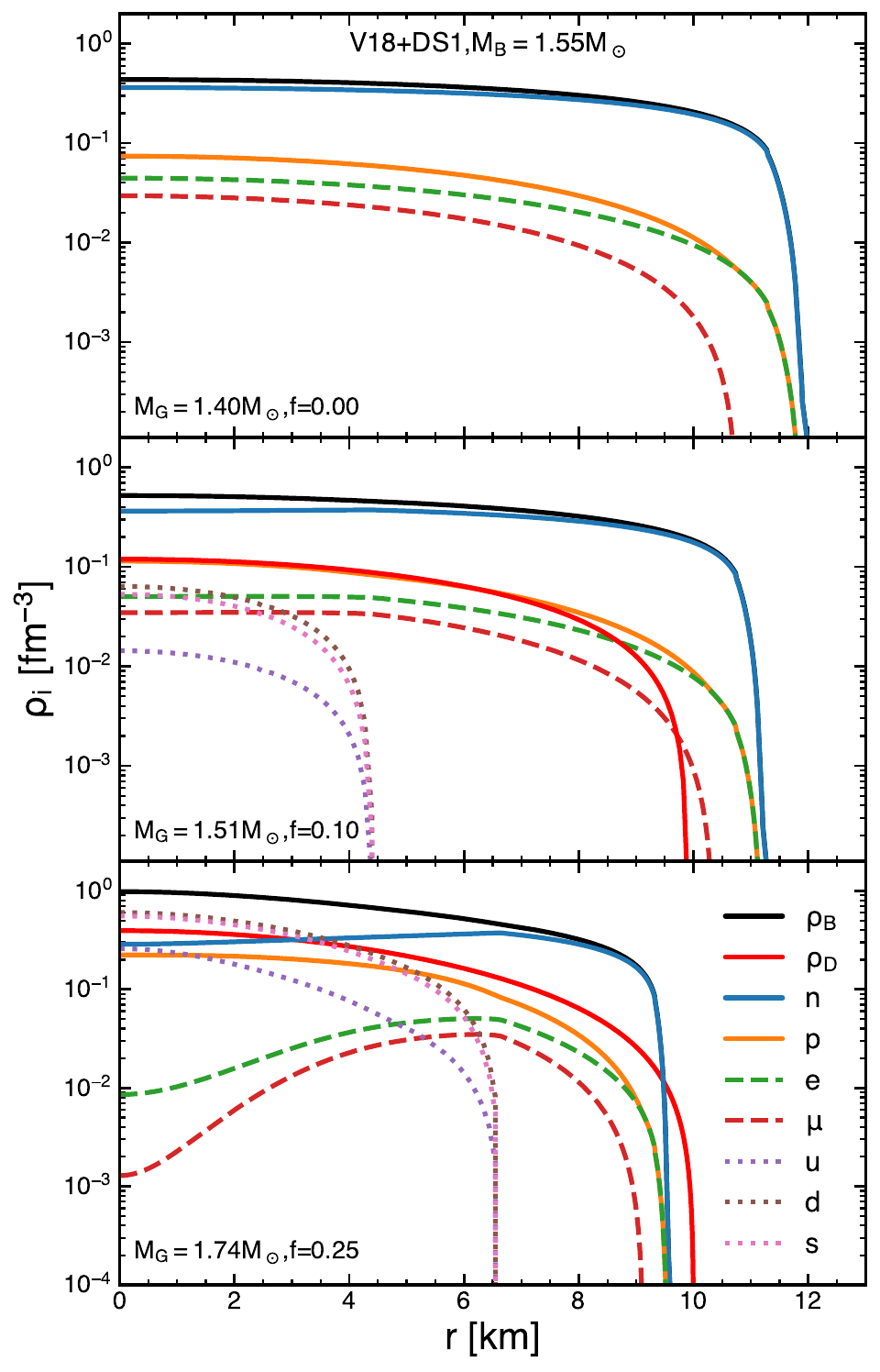}}
\vskip-4mm
\caption{
Profiles of particle-number densities of a DHS
of baryonic mass $M_B=1.55\ms$
at various DM fractions $f=0,0.1,0.25$
with the V18+DS1 EOS.
}
\label{f:nb}
\end{figure}

These oscillation equations are two coupled second-order homogeneous
differential equations for $\zeta_N$ and $\zeta_D$,
equivalent to a system of four first-order equations for
$\zeta_{N,D}$ and $\zeta_{N,D}'$.
For that purpose it is convenient to define
$\eta_i \equiv \hat{\Pi}_i \hat\zeta_i'$
and solve for $\zeta_i$ and $\eta_i$.
In order to do so,
one needs to impose boundary conditions.
The inner boundary conditions are
$\zeta_{N,D}(0) = 0$ and
$\eta_N(0) = \text{const.}$  
at the center of the star,
and the outer boundary conditions are
$\eta_N(R_N)=\eta_D(R_D)=0$
at the respective surfaces.
By solving these equations,
one obtains the central value $\eta_D(0)$
and the value of the squared fundamental radial oscillation frequency $\om^2$
of the star.
(We do not consider higher-order oscillation modes in this work). 
Stable DHS configurations are characterised by $\om^2>0$,
and an analysis will be given in Sec.~\ref{s:osc}.

\begin{figure*}[t]
\vskip-9mm
\centerline{\includegraphics[scale=0.65]{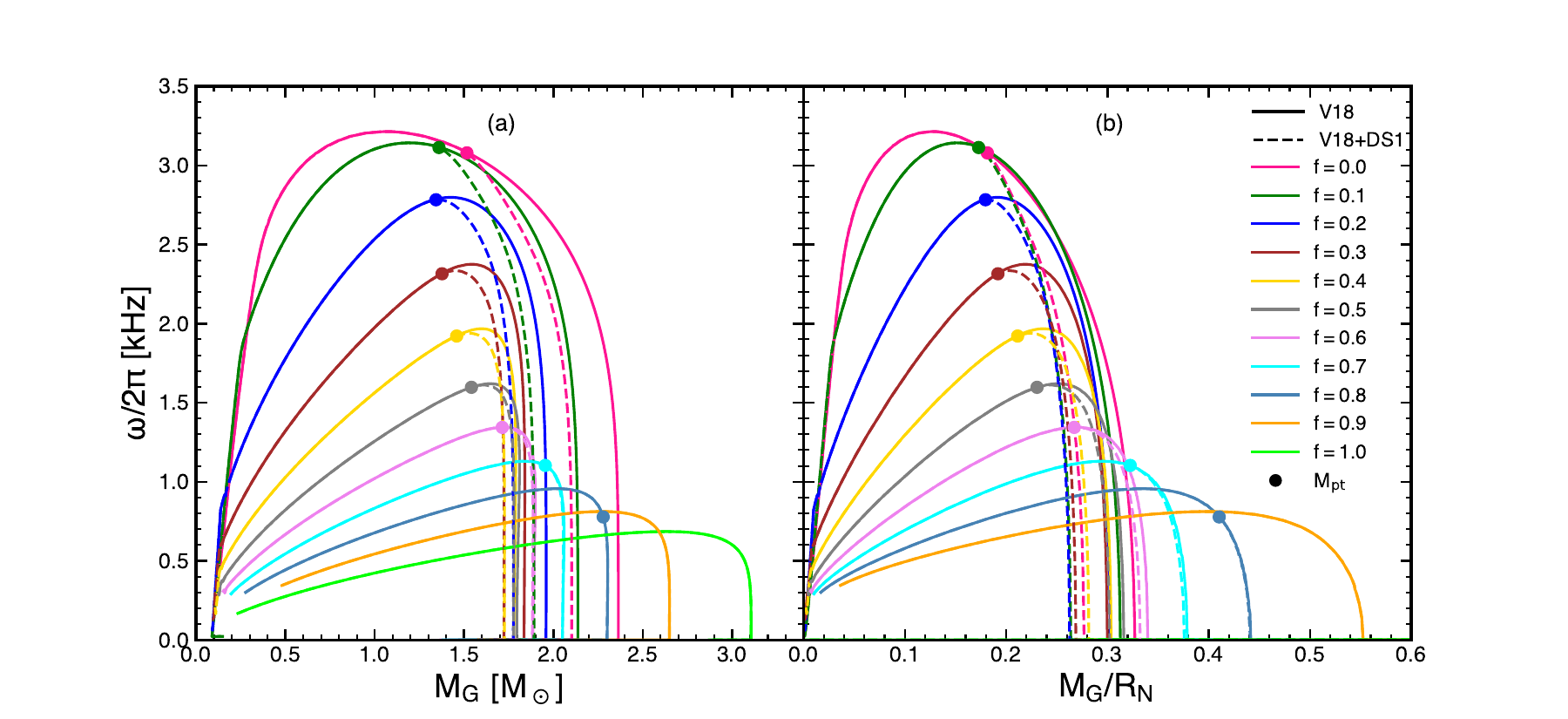}}
\vskip-4mm
\caption{
Fundamental radial oscillation frequency of DHSs
with V18 (solid) and V18+DS1 (dashed) EOSs
as a function of gravitational mass $M_G$
(a) and compactness $\beta=M_G/R_N$ (b)
for fixed $f=0,0.1,...,1$.
}
\label{f:fm}
\end{figure*}

\begin{figure*}[t]
\vskip-1mm
\centerline{\hskip0mm\includegraphics[scale=0.65]{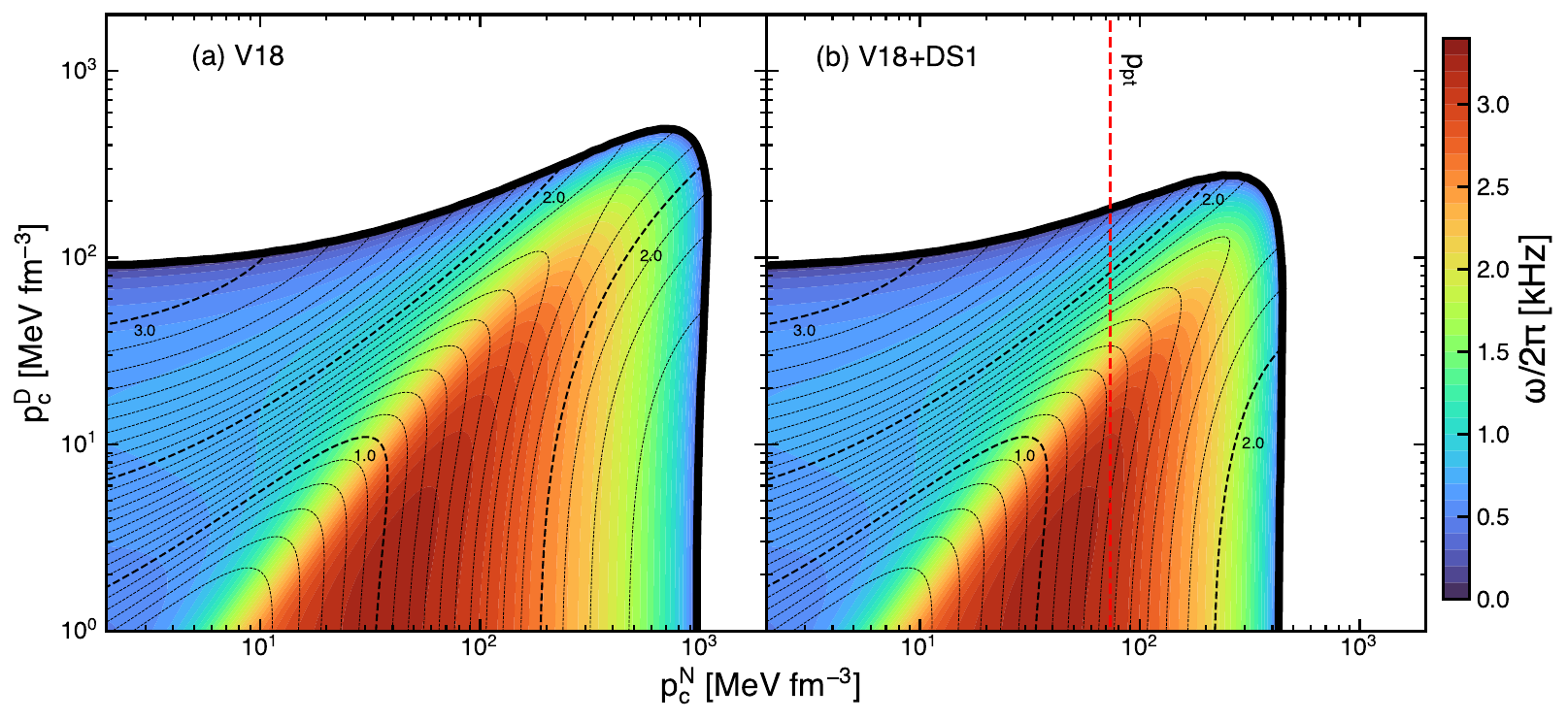}}
\vskip-4mm
\caption{
Heat plot of DHS radial oscillation frequencies
in the $(p_c^N,p_c^D)$ plane
for different EOS models.
Contours of $M_G/\ms$ (dashed black curves)
and the contour of the QM onset (vertical dashed red line in right panel)
are also shown.
}
\label{f:fpc}
\end{figure*}

\section{Results}
\label{s:res}

We first present the results of the structural TOV equations for DHSs,
and then the coupled structure + oscillation equations just introduced.

\subsection{Mass and radius of DHSs}

We begin in Fig.~\ref{f:mrf} with a plot of the mass-radius relations of
DHSs with varying DM fractions $f=0,0.1,0.2,...,1$,
for the pure nucleonic EOS and the three different hybrid EOSs
in the various panels.
Both gravitational mass $M_G$ (upper row)
and baryonic mass $M_B$ (lower row) are shown.
As exposed in previous work \cite{Liu24,Zhou25},
for the chosen value of the DM particle mass $\mu=1\gev$,
the effect of increasing DM fraction~$f$
on the $M_G(R)$ relation
is first (for $f\lesssim0.3$, DM-core regime)
a slight reduction of both $\mmax$ and radius,
and then (DM-halo regime) an increase of both up to $\mmax=3.11\ms$
of the pure dark star,
see also Fig.~\ref{f:mr}.

This general behavior depends only very weakly on the hybrid model,
as can also be seen in Fig.~\ref{f:mf},
which shows all possible stable DHS configurations in the $(f,M_G)$ plane.
In both Fig.~\ref{f:mrf} and Fig.~\ref{f:mf},
the solid black curves indicate the $\mmax$ configurations for fixed $f$,
and the dashed red curves the QM onset configurations.
In the latter case the qualitative behavior is the same as observed in
Figs.~\ref{f:eos} and \ref{f:mr}, namely,
the V18+DS1 model allows an earlier onset of QM than the other two models
(for any value of $f$).
This implies a wider range of DHS configurations
and consequently for the V18+DS1 EOS
HSs can reach significantly larger maximum masses at larger~$f$.
Note that at even larger $f$ beyond this point,
no hybrid configurations are possible,
due to the too strong reduction of the baryonic density.
The limiting configurations are indicated in both figures,
and depend on the quark model.
Fig.~\ref{f:mf} also shows the domains of DM-core and DM-halo stars.
For $\mu=1\gev$, most configurations with $f\gtrsim0.2$ are DM-halo stars.
For a more general analysis, see \cite{Liu24,Zhou25}.

\subsection{Effects of dark matter on hybrid stars}

The interplay of DM and QM inside a DHS has several interesting consequences.
For example, the presence of DM lowers the minimum mass of a HS
\cite{Lenzi23,Kumar25,Biesdorf25}:
In Fig.~\ref{f:mf}(b) it is seen that for the V18+DS1 model
the onset of QM is reduced from $M_G=1.45\ms$ at $f=0$
to about $1.31\ms$ at $f=0.19$.
The reason is the increase of the central baryon density due to the contraction
of a DM-core star caused by the added DM.

A related effect is the onset of QM in a NS \cite{Lenzi23,Biesdorf25}.
It is known that during the spin-down evolution of rotating NSs,
a hadron-quark phase transition could be induced in the stellar core.
This transition modifies the stellar moment of inertia
and thus the breaking index,
which might have observable consequences for pulsar timing signatures
\cite{Haensel16,Wei17,Prasad22}.
The accumulation of DM,
for instance in a dynamical formation scenario for DNSs,
where a preexisting NS is enveloped by a cloud of DM \cite{Gleason22},
may thus act as an alternative trigger for such a phase transition in a NS.
This process may produce potentially observable pulsar signals.
In addition to those,
once QM appears in the core,
fast neutrino emission through the quark direct Urca process
could become operative.
This can strongly affect the thermal evolution of the star
and modify the observable photon emission from the surface
\cite{Iwamoto82,Yakovlev01}.

This is better seen in the $(f,M_B)$ diagram Fig.~\ref{f:mbf},
for the V18+DS1 EOS.
In the specific DM accretion scenario,
the baryonic mass is approximately constant,
and the stellar configurations follow the corresponding trajectories,
some of which
[the ones belonging to initial gravitational mass
$M_0=M_G(f=0)=1.0,1.2,1.4,1.6\,\ms$]
are shown as horizontal green lines.
It is seen that below the QM onset mass $M_0=1.45\ms$ of a normal NS,
the increase of $f$ (addition of DM)
achieves QM onset at the critical fraction $f_\text{pt}$ of the added DM,
for example
$f_\text{pt}=0.02$ for $M_0=1.4\ms$,
$f_\text{pt}=0.11$ for $M_0=1.2\ms$.

Fig.~\ref{f:nb} illustrates details of the transition to a HS
for $M_0=1.4\ms$ ($M_B=1.55\ms$)
with the V18+DS1 EOS
by means of the density profiles
$\rho_i(r),\ i=D,B,n,p,e,\mu,u,d,s$
at $f=0,\;0.1,\;f_\text{max}=0.25$.
The transition proceeds mainly in the DM-core regime,
as seen in Fig.~\ref{f:mbf},
and the total radius $R$ is shrinking
from $12.3\km$ ($f=0$) to $10.0\km$ ($f=0.25$).
QM onset occurs at $f=0.02$ such that at $f=0.1$ an extended QM-admixed core
is already present,
although NM is still dominant.
This QM core grows to about $R_Q=6.5\km$ at the maximum possible $f=0.25$,
where collapse to a black hole would occur.
The gravitational mass is $1.74\ms$ at this point.
The physical reason for the appearance of QM
is the rise of the (central) baryon density $\rho_B$
(solid black curves)
with increasing $f$,
as clearly seen in the figure,
due to the presence of the background DM-core density $\rho_D$
(solid red curves).

\subsection{DHS radial oscillations}
\label{s:osc}

Solving the oscillation equations (\ref{e:osc})
for a given stellar configuration,
namely a solution of the structure equations (\ref{e:tovd}-\ref{e:tovnu}),
one obtains the squared fundamental radial oscillation frequency $\om^2$,
which indicates stability if positive.
The result $\om(M_G)$ is shown in Fig.~\ref{f:fm}(a)
for all $(M_G,f)$ configurations discussed before
in Fig.~\ref{f:mrf},
for both V18 and V18+DS1 EOS.
In all cases the frequency remains positive
up to the maximum mass along the fixed-$f$ curves,
where it vanishes.

Further exotic stable configurations as studied in some recent works
\cite{Hippert23,Barbat24,Zhen24,Pitz25,Biesdorf25} are not excluded,
but disregarded in the present work.
In fact it has been shown in stability analyses of the two-fluid model
\cite{Henriques90b,Leung12,Valdez13,Kain20,Kain21,Leung22,Gleason22,
Hippert23,Barbat24,Zhen24,Pitz25,Biesdorf25}
that the limits of stellar stability occur at the
maximum of $M$ along fixed-particle-number contours,
and this has been demonstrated \cite{Leung12,Kain20,Leung22,Zhen24}
to coincide quite accurately with the maximum of the $M(R)$
relations for fixed DM fraction $f$,
just as in the limit of ordinary NSs.
Configurations with larger central densities (smaller radius)
are unstable with respect to induced radial oscillations,
as verified in the above references.

In Fig.~\ref{f:fm}(a)
one notes that the frequency of pure NSs might exceed 3\,kHz
over a wide mass range,
as usually found for ordinary NSs with any realistic EOS
\cite{Glass83,Gondek97,Kokkotas01,Gupta02,Sahu02,Brillante14,
Sagun20,Sotani21,Sun21,Hong22,Zhen24,Ghosh25},  
while an increasing DM fraction reduces it continuously
to less than 1\,kHz for the pure dark star.
This effect is caused by the coupling of the two co-oscillating fluids
and the increased compactness of the baryonic matter
for not too large~$f$,
see, e.g., Fig.~\ref{f:nb}.
Therefore, observation of a compact object with moderate mass $\sim1.4\ms$
and very low radial frequency~\hbox{$\sim1-2\,$kHz}
might be a signal for a substantial DM content.

In fact it has been shown \cite{Sun21,Ghosh25} that for HSs
the dependence of $\om$ on the stellar compactness $\beta=M_G/R_N$
(note that $R_N$ is the observable optical radius)
depends only very weakly on the nuclear EOS, i.e.,
$\om(\beta)$ is a nearly universal relation for hadronic NSs,
but not for DM-admixed stars.
This is demonstrated in Fig.~\ref{f:fm}(b),
showing the strong reduction of the frequency by added DM
for $\beta\lesssim0.2$,
corresponding to the moderate-mass range in panel (a).
On the other hand,
also very large values of $\beta\gtrsim0.4$ might indicate DM,
due to the reduction of the optical radius $R_N$.

Instead of using $M_G$ and $f$ as independent variables
characterizing the stellar configurations,
the stability analysis is often
\cite{Gleason22,Barbat24,Pitz25,Kumar25}
carried out by using instead the central pressures
$p_c^{N,D} \equiv p_{N,D}(r=0)$.
Fig.~\ref{f:fpc} shows the heat map of the radial oscillation frequency
(color scale)
in the $(p_c^N,p_c^D)$ plane
for both V18 (a) and V18+DS1 (b) EOSs.
Note that the two panels differ  only in the
$p_c^N>p_\text{pt}=73\mfm$ domain,
to the right of the corresponding contour in panel (b).
The thick solid black curve marks the condition $\om=0$,
which defines the boundary of radial instability.
Stellar configurations located beyond this curve are unstable.
The thin dashed black curves are the contours of $M_G$,
and it can be seen that the stability limit $\om=0$
does not coincide with $\mmax$ along fixed-$p_c^i$ curves.

As shown in both Fig.~\ref{f:fm} and \ref{f:fpc},
the radial oscillation frequencies are strongly affected by DM.
The maximum frequency occurs at vanishing DM content
for both pure NSs and HSs,
and the introduction of DM leads to a systematic reduction.
This behavior can be understood as a consequence of the competition
between the two coupled fluid components:
It is seen that the radial oscillations of pure dark stars
($f=1,p_c^N=0$)
have much lower frequencies than those of pure NSs.
\OFF{
The frequency of the radial oscillations is influenced
by both the stellar radius and the speed of sound,
approximately being proportional to $\frac{1}{R}$ and $c_s$.
The presence of DM modifies both the mass-radius relation
and the internal sound-speed profile of the star.
}
Therefore,
the inclusion of DM effectively reduces the overall frequency for DHSs.
This is a consequence of our choice of $\mu=1\gev$ for DM.
By contrast, as discussed in \cite{Kain21,Kumar25},
for a sufficiently soft DM EOS
(such as $\mu = 2\gev$),
by the same mechanism
the maximum frequency appears at relatively large DM central pressures
and fractions,
due to the small radius.
For mirror DM models,
the frequency distribution is symmetric in the $(p_c^N,p_c^D)$ plane.

Another related result is that two-fluid stars
can reach higher central pressures and densities
compared to single-fluid configurations.
For example, in Fig.~\ref{f:fpc}(a),
the maximum $p_c^D$ increases from $91\mfm$ to $489\mfm$,
while $p_c^N$ increases slightly from $975\mfm$ to $1080\mfm$.
As said, this effect can shift the QM onset to lower masses.
In Fig.~\ref{f:fpc}(b),
for the fixed QM onset pressure $p^N_c=p_\text{pt}=73\mfm$
(vertical dashed red line),
the gravitational mass first decreases and then increases
as the DM content grows,
in line with Fig.~\ref{f:mf}(b).
Correspondingly, the oscillation frequency decreases
and eventually vanishes at the stability boundary.

\section{Summary}
\label{s:sum}

We have examined the interplay of DM and QM in DHSs
modeled by a realistic BHF nuclear EOS,
a representative fermionic DM EOS,
and three QM models respecting the important constraint on the NS maximum mass
in a Gibbs construction involving NM and QM.
We studied the stable solutions of the corresponding
coupled structure and oscillation equations.

The DM degree of freedom in terms of a prescribed DM fraction $f$
influences the internal stellar structure,
in particular the possible presence of QM.
We exposed and confirmed several related effects:
(i) the minimum mass of a HS is lowered by DM,
due to an induced increase of the baryon density;
(ii) this facilitates a possible onset of QM formation in a DM-accreting star;
(iii) radial oscillation frequencies are strongly affected by DM,
which decreases them by the coupling with the NM fluid,
and might thus indicate the presence of DM in a compact object.

These are qualitative effects that all depend on the details of
the three relevant EOSs for NM, QM, and DM,
of which the latter two are currently especially uncertain.
We mention in particular the dependence on the DM particle mass,
which has not been explored in this work,
but in previous ones \cite{Liu24,Zhou25}.
Future observational data might help to reduce our pertaining ignorance.

\section*{Acknowledgments}

This work is sponsored by the National Natural Science Foundation of China
under Grant No.~12205260.

\def\aap{Astron. Astrophys.}
\def\aap{A\&A}
\def\apjl{Astrophys. J. Lett.}
\def\apjs{ApJS}
\def\araa{Annu. Rev. Astron. Astrophys.}
\def\epja{EPJA}
\def\epjc{EPJC}
\def\jcap{J. Cosm. Astropart. Phys.}
\def\jcap{JCAP}
\def\jpg{J. Phys. G}
\def\mnras{Mon. Not. R. Astron. Soc.}
\def\mnras{MNRAS}
\def\npb{Nucl. Phys. B}
\def\physrep{Phys. Rep.}
\def\plb{Phys. Lett. B}
\def\ppnp{Prog. Part. Nucl. Phys.}
\def\rpp{Rep. Prog. Phys.}
\bibliographystyle{apsrev4-1}
\bibliography{dmhs}

\end{document}